\documentclass[aps,pra,twocolumn,showpacs,superscriptaddress,letterpaper,floatfix]{revtex4}
\usepackage{graphicx}
\usepackage{epsfig}

\usepackage{longtable}
\usepackage{amssymb}
\usepackage{amsthm}
\usepackage{amsfonts}
\usepackage[centertags]{amsmath}
\usepackage{color}
\usepackage{sidecap}
\usepackage{pst-all}
\usepackage{pstricks}
\usepackage[latin1]{inputenc}
\usepackage[english]{babel}
\usepackage{soul}
\usepackage{pifont} 

\newcommand{\coon}{(Color online) }

\newcommand{\expec}[1]{\langle#1\rangle}  

\newcommand{\p}{{\phantom{0}}}
\newcommand{\beq}{\begin{equation}}
\newcommand{\eeq}{\end{equation}}
\newcommand{\irm}{{\rm i}}
\newcommand{\e}{{\rm e}}

\begin{document}

\title{Coherent control of broadband vacuum squeezing}

\author{Simon Chelkowski}
\affiliation{Max-Planck-Institut f\"ur Gravitationsphysik
(Albert-Einstein-Institut) and\\ Institut f\"ur
Gravitationsphysik, Leibniz Universit\"at Hannover, Callinstr.\
38, 30167 Hannover, Germany}

\author{Henning Vahlbruch}
\affiliation{Max-Planck-Institut f\"ur Gravitationsphysik
(Albert-Einstein-Institut) and\\ Institut f\"ur
Gravitationsphysik, Leibniz Universit\"at Hannover, Callinstr.\
38, 30167 Hannover, Germany}

\author{Karsten Danzmann}
\affiliation{Max-Planck-Institut f\"ur Gravitationsphysik
(Albert-Einstein-Institut) and\\ Institut f\"ur
Gravitationsphysik, Leibniz Universit\"at Hannover, Callinstr.\
38, 30167 Hannover, Germany}

\author{Roman Schnabel}
\affiliation{Max-Planck-Institut f\"ur Gravitationsphysik
(Albert-Einstein-Institut) and\\ Institut f\"ur
Gravitationsphysik, Leibniz Universit\"at Hannover, Callinstr.\
38, 30167 Hannover, Germany}

\date{\today}

\begin{abstract}
We present the observation of optical fields carrying squeezed
vacuum states at sideband frequencies from 10\,Hz to above
35\,MHz. The field was generated with type-I optical parametric
oscillation below threshold at 1064\,nm. A coherent, unbalanced
classical modulation field at 40\,MHz enabled the generation of
error signals for stable phase control of the squeezed vacuum
field with respect to a strong local oscillator. Broadband
squeezing of approximately $-4$\,dB was measured with balanced
homodyne detection.  The spectrum of the squeezed field allows a
quantum noise reduction of ground-based gravitational wave
detectors over their full detection band, regardless of whether
homodyne readout or radio-frequency heterodyne readout is used.
\end{abstract}


\maketitle

\section{Introduction}

Quantum noise is one of the limiting noise sources in laser
interferometers and appears as an uncertainty of the light field's
quadratures, which carry the signal of the interferometric
measurement. For coherent laser radiation the quantum noise is
minimal, and symmetrically distributed among pairs of noncommuting
field quadratures. The finite, non-zero value of this so-called
vacuum noise is a manifestation of Heisenberg's uncertainty
relation (HUR). However, for certain nonclassical states of light
the quantum noise can be asymmetrically distributed among field
quadratures, such as amplitude and phase quadratures. In the case
of squeezed states \cite{Wal83}, the quantum noise of one
quadrature is reduced below vacuum noise, whereas the quantum
noise in the orthogonal quadrature is increased without violating
the HUR. Since an interferometer measures a certain, single
quadrature of the field, an appropriately squeezed field can
improve the signal-to-noise ratio in a quantum-noise-limited
interferometer.

The application of squeezed states in laser interferometers was
first proposed by Caves  \cite{Cav81} in 1981. Motivated by the
challenging effort at the direct observation of gravitational
waves \cite{Thorne87}, Caves suggested injecting squeezed vacuum
states of light into the dark signal port of interferometric
gravitational wave detectors.  The goal of that proposal was the
reduction of the vacuum noise of the interferometer's readout
laser beam, which is often called shot noise. Two years later
Unruh~\cite{Unruh82} realized that squeezed light can be used to
correlate interferometer shot noise and radiation pressure noise
(back-action noise) in such a way that the so-called standard
quantum limit can be broken, and a quantum nondemolition
measurement on the mirror test mass position can be performed. For
an overview we refer the reader to Ref.\,\cite{KLMTV01}. The
theoretical analysis of Harms {\it et al.}\,\cite{HCCFVDS03}
further motivated research on squeezed states. They found that
advanced interferometer recycling techniques \cite{Mee88} that
also aim for an improvement of the signal-to-shot-noise ratio are
also fully compatible with squeezed-field injection.

The first observation of squeezed states was done by Slusher {\it
et al.~}\cite{SHYMV85} in 1985. Since then different techniques
for the generation of squeezed light have evolved. One of the most
successful approaches to squeezed-light generation is optical
parametric oscillation (OPO). Hence common materials like
MgO:LiNbO$_{3}$ can be used to produce broadband squeezing at the
carrier wavelength of today's gravitational wave (GW) detectors
(1064\,nm). In the future, various recycling techniques as well as
the most powerful single-mode lasers available will be used to
reduce the quantum noise in GW detectors. It is generally expected
that the interferometer sensitivities will be limited by shot
noise in the upper audio band  and by radiation pressure noise in
the lower audio band \cite{Shoemaeker03}. At intermediate
frequencies, both quantum noise and thermal noise \cite{BGV99} are
expected to dominate the overall noise floor. Therefore squeezing
of quantum noise indeed offers a further increase of GW detector
sensitivities.

Gravitational wave detectors require a broad\-band squeezed field
in the detection band from about 10\,Hz to 10\,kHz. If a
radio-frequency (rf) heterodyne readout is used, squeezing in the
band of 10\,kHz around twice the rf-phase modulation frequency is
also required \cite{GLe87}. Furthermore, GW detectors utilize
recycling cavities, implying that the orientation of the squeezing
ellipse needs to be designed for every sideband frequency. The
transformation from frequency-independent squeezing to optimized
frequency-dependent squeezing can be performed by optical filter
cavities, as proposed in \cite{KLMTV01} and demonstrated in
\cite{CVHFLDS05}. The combination of squeezed-field injection and
optimized orientations of squeezing ellipses, as well as power
recycling and signal recycling of interferometers, has been
demonstrated in \cite{VCHFDS05,VCHFDS06CQG}. Squeezed states at
audio frequencies have been demonstrated recently \cite{MGBWGML04,
MMGLGGMM05,VCHFDS06}.

Applications of squeezed states generally require active phase
control with respect to the local oscillator field of the readout
scheme. Controlling the squeezed fields is indeed the basic
problem for squeezed-field applications in GW detectors. Common
control schemes rely on the injection of a weak, phase-modulated
seed field at the carrier frequency into the OPO, thereby turning
the device into an optical parametric amplifier (OPA). It has been
shown that even the lowest carrier powers introduce too large
amounts of classical laser noise at audio frequencies, and
squeezing can no longer be achieved \cite{MGBWGML04}. On the other
hand, phase modulation sidebands are not present in a pure vacuum
field. For this reason a coherent control field could not be
created in \cite{MGBWGML04} for either the squeezed-field carrier
frequency or the relationship between the squeezed quadrature
angle and local oscillator. The quadrature angle was locked
instead using so-called noise locking, whose stability was found
to be significantly less than what can be achieved with coherent
modulation locking \cite{MMGLGGMM05} as used in GW
interferometers.

In this paper we report on the generation and coherent control of
broadband squeezing from subaudio frequencies up to radio
frequencies. The coherent control scheme that was first used in
\cite{VCHFDS06} is presented in detail.

\section{Control Scheme} \label{sec2}

\begin{figure*}[t]
\centerline{\includegraphics[width=17.2cm]{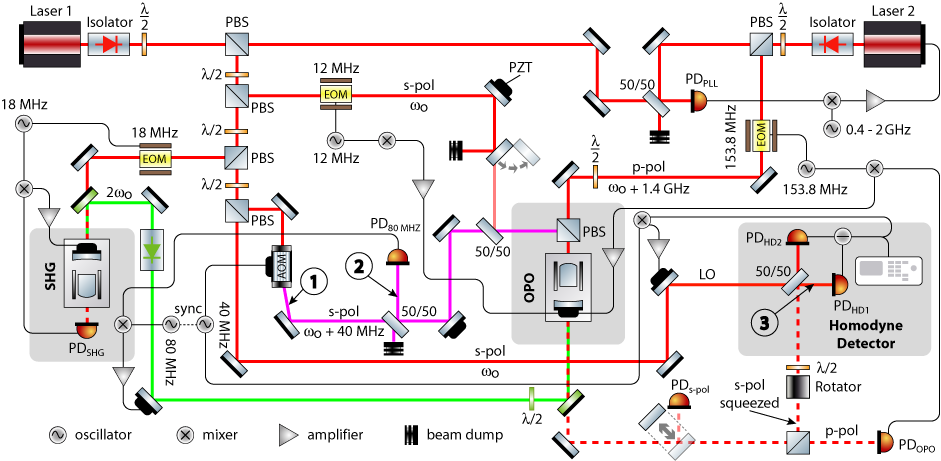}} 
\vspace{0mm} \caption{\coon  Schematic of the experiment.
Generation and full coherent control of a broadband squeezed
vacuum field at 1064\,nm was achieved utilizing two independent
but phase-locked laser sources. Laser 1 provided the main carrier
frequency of homodyning local oscillator ($\omega_{0}$). It also
provided the quadrature control field (QCF), which was frequency
shifted through an acousto-optical modulator (AOM), and the
optical parametric oscillator (OPO) pump field, which was produced
through second harmonic generation (SHG). Laser 2 provided another
frequency-shifted control field for locking the OPO cavity length.
PBS: polarizing beam splitter;  DC: dichroic mirror; LO: local
oscillator, PD: photo diode; EOM: electro-optical modulator, PZT:
piezoelectric transducer.}
  \label{experiment}
\end{figure*}

\begin{figure}[t]
\centerline{\includegraphics[width=8cm]{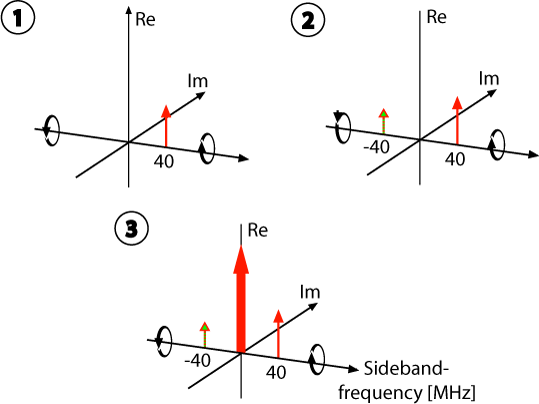}} 
\vspace{0mm} \caption{\coon Complex optical field amplitudes at
three different locations in the experiment, which are marked in
Fig.~\ref{experiment}.}
  \label{sidebands}
\end{figure}

All current interferometric gravitational wave detectors are
Michelson interferometers operating close to a dark fringe at the
signal output port. The optical field at the output port consists
of a local oscillator field that beats with modulation sideband
fields at frequencies $\Omega_s$ generated by gravitational waves
and (quantum) noise. In homodyne detection, the local oscillator
has the same optical frequency $\omega_{0}$ as the main
interferometer laser field. In heterodyne detection the local
oscillator consists of a combination of upper and lower modulation
sideband fields at frequencies $\omega_{0}\pm\omega_m$. Both
detection schemes provide eigenvalues of the time-dependent
quadrature operator $\hat q_\theta (\Omega_s, \Delta\Omega, t)$,
where $\Delta\Omega$ is the resolution bandwidth (RBW) and
$\theta$ the quadrature angle. The angle $\theta$ might be chosen
to select the quadrature with the optimum signal-to-noise ratio.
In the following we will refer to the amplitude quadrature
($\theta=0$) and the phase quadrature ($\theta=90^{\circ}$) with
the subscripts 1 and 2.
For vacuum fields, the variances of the quadrature operators are
typically normalized  to unity and the Heisenberg uncertainty
relation sets the following lower bound for the product of the
quadrature variances:
\begin{equation} \label{hur}
 \Delta \!  ^{2} \hat q_1(\Omega_s, \Delta\Omega, t) \times \Delta \!  ^{2} \hat q_2(\Omega_s, \Delta\Omega, t) \ge 1\,.
\end{equation}
For a broadband amplitude-squeezed field, $ \Delta \!  ^{2} \hat
q_1(\Omega_s, \Delta\Omega, t)$ is always below the unity vacuum
noise reference for all sideband frequencies within the squeezing
band. The HUR requires that in this case $ \Delta \!  ^{2} \hat
q_2(\Omega_s, \Delta\Omega, t)$ is greater than unity by a factor
of at least the inverse of the squeezed quadrature variance. A
squeezed vacuum field is said to be pure if, for all sideband
frequencies in a certain band, the equals sign holds in
Eq.\,(\ref{hur}). If a pure, squeezed vacuum field senses optical
loss due to absorption or scattering, the squeezed field gets
mixed with the (ordinary) vacuum field. In that case the equals
sign in Eq.\,(\ref{hur}) is no longer realized; however, we still
speak of a squeezed vacuum field or just vacuum squeezing.

For the application of a squeezed vacuum field in an
interferometer, the squeezed quadrature needs to be matched to the
interferometer readout quadrature. To achieve this goal the OPO
cavity needs to be length controlled to resonate for the carrier
frequency $\omega_{0}$. Furthermore, the wave front of the
second-harmonic OPO pump field has to be phase controlled with
respect to the interferometer readout field.  Note that the phase
of the OPO pump field determines the phase of the squeezed field.
Both control requirements mentioned can easily be realized if a
radio-frequency phase-modulated field at carrier frequency, that
is sent through the OPO cavity, can be utilized. If such a field
cannot be applied, for example because its noise prevents the
observation of squeezed states, coherent control is much more
difficult to achieve. In this section we discuss in detail the
coherent control scheme that was first used in \cite{VCHFDS06}.

Our scheme uses control fields that are coherent with the squeezed
field without interfering with it. The latter ensures that noise
from the control fields does not deteriorate the nonclassical
performance of the squeezed field. Altogether two coherent control
fields are required. Both are frequency shifted against the
carrier frequency $\omega_{0}$. A coherent, frequency-shifted
field can be generated by an independent but phase-locked laser
source or by an acousto-optical modulator (AOM) acting on a tapped
laser beam from the same source. The first frequency-shifted
control field enables length control of the OPO cavity. It carries
radio-frequency phase modulation sidebands, is orthogonally
polarized with respect to the squeezed vacuum field, and is
injected into the OPO cavity. The frequency shift should exactly
compensate the birefringence of the nonlinear crystal such that
both control field and squeezed field simultaneously resonate in
the OPO cavity. The frequency shift as well as the orthogonal
polarization prevents the interference of the squeezed vacuum
field and the control field.

The control of the quadrature angle of the squeezed vacuum with
respect to a local oscillator or an interferometer carrier field
is more challenging. This request is achieved by the second
control field (quadrature control field, QCF) which is also
injected into the OPO cavity. This field \emph{does} sense the OPO
nonlinearity but is frequency shifted against the vacuum squeezed
mode. The QCF allows the generation of two individual error
signals for two different servo control loops. The first error
signal is used to control the relative phase between the QCF and
the OPO pump field. The second error signal is used to control the
relative phase between the OPO pump field and the local oscillator
field of the interferometer, or rather the homodyne detector. The
combination of both error signals provides the means to stabilize
the quadrature of the squeezed vacuum field with respect to a
local oscillator.

In the following we show that the required error signals can be
gathered from the QCF leaving the OPO cavity, and from the
interference of the QCF with the local oscillator field at the
homodyne detector. We label those two error signals
$S\mathrm{^{QCF-P}_{err}}$ and $S\mathrm{_{err}^{QCF-LO}}$,
respectively, and first derive an expression for the
parametrically amplified quadrature control field QCF.
Before parametric amplification the QCF at optical frequency
$\omega_0 + \Omega$ represents a single sideband field with
respect to the carrier frequency $\omega_0$ (see \ding{192} in
Figs.~\ref{experiment} and \ref{sidebands}). In the following we
describe this field by the real-valued amplitude $\alpha_\Omega$.
The expectation values of the annihilation operators of the upper
and lower sideband fields at frequencies $\omega_0 \pm \Omega$ may
then be written as follows:

\begin{equation}
\begin{split}
\expec{\hat a_+} & \equiv \expec{\hat a(\omega_0+\Omega)}=\alpha_\Omega\,,\\
\expec{\hat a_-} & \equiv \expec{\hat a(\omega_0-\Omega)}=0 \,.
\end{split}
\end{equation}
The quadrature amplitudes  \cite{CSc85} are given by
\begin{equation}
\hat a_1^\p=\dfrac{1}{\sqrt{2}}(\hat a_+^\p+\hat
a_-^\dagger)\,,\quad \hat a_2^\p=\dfrac{1}{\irm\sqrt{2}}(\hat
a_+^\p-\hat a_-^\dagger)\,. \label{quadratureamplitudes}
\end{equation}
Here all quantities are defined for discrete frequencies. This
simplifies our description and is reasonable because the bandwidth
of the error signals is small compared to $\Omega$.

The OPO acts on these quadrature amplitudes in different ways. If
it amplifies the phase quadrature then it deamplifies the
amplitude quadrature, and vice versa. Mathematically, this effect
of amplification and deamplification of the quadratures
$\mathbf{\bar a}\!=\! \left(
\begin{array}{c}
  \!  \hat a_1\!\\
  \!  \hat a_2\!
\end{array}\right)$
can be described with the use of the squeezing operator
$S(r,\phi)=\textrm{exp}[r(\hat a_+\hat a_-\e^{-2\irm\phi}-\hat
a_+^\dagger\hat a_-^\dagger\e^{2\irm\phi})]$ with squeezing factor
$r$ and squeezing angle $\phi$ (see \cite{CSc85}). The resulting
squeezed quadrature vector $\mathbf{\overline{b}}$ is given by:
\begin{equation}
    \begin{split}
\mathbf{\overline{b}}&= \left(
\begin{array}{c}
    \hat b_1(\Omega)\\
    \hat b_2(\Omega)
\end{array}\right) =S(r,\phi)\mathbf{\bar a}S^\dag(r,\phi)
\\&= \left(
\begin{array}{c c}
\cosh(r)+\sinh(r) C_{2\phi} & \sinh(r) S_{2\phi} \\
\sinh(r) S_{2\phi} & \cosh(r)-\sinh(r) C_{2\phi}
\end{array} \right) \mathbf{\bar a}\,,
\end{split}
\label{Ebar}
\end{equation}
where $C_{2\phi}=\cos(2\phi)$ and $S_{2\phi}=\sin(2\phi)$.

The expectation values of the new squeezed quadrature amplitudes
$\hat b_1(\Omega)$ and $\hat b_2(\Omega)$ take the following form
\begin{equation}
\begin{split}
\expec{\hat b_1(\Omega)}&= -\frac{\irm
\alpha_\Omega}{{\sqrt{2}}}S_{2\phi}\sinh (r) +
\frac{\alpha_\Omega}{{\sqrt{2}}}\left[ \cosh(r)\!+\!C_{2\phi}\sinh
(r) \right]\,,\\ \expec{\hat b_2(\Omega)}&=
\frac{\alpha_\Omega}{{\sqrt{2}}}S_{2\phi}\sinh(r) - \frac{\irm
\alpha_\Omega}{{\sqrt{2}}}\left[\cosh (r)\!-\!C_{2\phi}\sinh(r)
\right]\,.
\end{split}
\label{squeezedquadratureamplitudes}
\end{equation}
To derive the corresponding electrical field
\begin{align}
    E^{\textrm{QCF}}(t) &\propto \expec{\hat b^{(+)}(t) + \hat b^{(-)}(t)}
    \intertext{with}
    \hat b^{(\pm)}(t)&\equiv\frac{1}{2}\left[\hat b_1(t)\pm \irm \hat b_2(t)\right]\e^{\mp \irm\omega_{0}
    t}
\end{align}
we need the Fourier transformations of $\hat b_1(\Omega)$ and
$\hat b_2(\Omega)$. Since we consider a single frequency we obtain
\begin{equation}
\begin{split}
    \hat b_1(t) &=  \hat b_1(\Omega) \e^{-\irm\Omega t} + \hat b_1^*(\Omega) \e^{\irm\Omega t} \,,\\
    \hat b_2(t) &=  \hat b_2(\Omega) \e^{-\irm\Omega t} + \hat b_2^*(\Omega) \e^{\irm\Omega t} \,.
\end{split}
\end{equation}
By choosing $\exp{(r)}=\sqrt{g}$ we simplify our expression and
find for the outgoing QCF from the OPO
\begin{align}\label{E(t)}
E^{\textrm{QCF}}(t) &\propto \frac{1 +
g}{\sqrt{2\,g}}\,\alpha_\Omega \,\cos (\omega_{0}\,t  +
\Omega\,t)\nonumber\\
&\p\quad -\frac{1 - g}{\sqrt{2\,g}}\,\alpha_\Omega \,
     \cos (\omega_{0}\,t  - \Omega\,t  - 2\,\phi)\,.
\end{align}
This is the desired expression for the parametrically amplified
QCF, and forms the basis for the following derivation of the two
error signals $S\mathrm{^{QCF-P}_{err}}$ and
$S\mathrm{_{err}^{QCF-LO}}$. One can easily see in
Eq.\,(\ref{E(t)}) that $E^{\textrm{QCF}}(t)$ is composed of two
sidebands that are equally separated by $\Omega$ from the carrier
frequency $\omega_{0}$. This is also illustrated in the sideband
scheme \ding{193} in Fig.~\ref{sidebands}. The quadrature where
these two sidebands beat with each other can be chosen using the
squeezing angle $\phi$. If one uses a squeezing angle of $\phi=0$
the following electrical field is found:
\begin{equation}
\begin{split}
E^{\textrm{QCF}}(t)_{\phi=0}&\propto
\frac{1}{\sqrt{2}}\left(\sqrt{g} \cos(\Omega t)
-\irm\frac{1}{\sqrt{g}} \sin(\Omega t)\right)\\&\p\quad\quad\quad\quad\quad \times a_\Omega(\Omega) \e^{- \irm\omega_{0} t}+ \textrm{c.c.}\,,\\
\end{split}
\end{equation}
which has also been provided in \cite{VCHFDS06}.

Detection of the outgoing field $E^{\textrm{QCF}}(t)$ from the OPO
with a single photo diode results in the following photocurrent:
\begin{equation}
\begin{split}
I^{\textrm{QCF}}&\propto\frac{{\alpha_\Omega}^2}{2\,g}\,\left[\left(
1
+ g \right) \, \cos \left( \omega_{0}\,t  + \Omega\,t \right)\right.\\
&\p\quad-\left.\left( 1 - g
\right)\,\cos\left(\omega_{0}\,t-\Omega\,t-2\,\phi\right)\right]^2\,.\\
\end{split}\label{I_AOM}
\end{equation}
Demodulating $I^{\textrm{QCF}}$ with frequency $2\Omega$ and
subsequent low-pass filtering provide the error signal
$S\mathrm{^{QCF-P}_{err}}$ for the relative phase between the
second-harmonic pump field and the QCF, given in terms of the
squeezing angle $\phi$

With an appropriate demodulation phase one obtains {the sinusoidal
error signal}
\begin{equation}
 S\mathrm{^{QCF-P}_{err}}\varpropto\frac{\left( -1 + g^2 \right) \,{\alpha }^2\,\sin (2\,\phi )}{4\,g} \,.
 \label{Err_AOM}
\end{equation}
Now that we are able to stabilize $\phi$ with respect to the QCF,
the first step to a complete coherent control of a squeezed vacuum
generated by an OPO is satisfied.

In a second step the phase $\Phi$ between the second-harmonic pump
field and the local oscillator needs to be controlled. The error
signal $S\mathrm{_{err}^{QCF-LO}}$ is generated from the
difference current of the two homodyne photodiodes
PD$_\textrm{HD1,2}$. {Overlapping the local oscillator field
$E^{\textrm{LO}} \propto \alpha^{\textrm{LO}} \e^{-\irm\omega_{0}
t} \e^{-\irm\Phi} + \textrm{c.c.}$ with the outgoing QCF
$E^{\textrm{QCF}}(t)$ from the OPO at the {\textrm{LO}} homodyne
beam splitter results in two homodyne detector fields
$E^{\textrm{HD1}}$ and $E^{\textrm{HD2}}$ which are individually
detected with a single photodiode. The complex field amplitudes of
one of the homodyne detector fields can be seen graphically in
Fig.~\ref{sidebands},\ding{194}. Mathematically they are given by}
\begin{align}
   E^{\textrm{HD1}} & = \nonumber \frac{1}{\sqrt{2}}\left[E^{\textrm{LO}} + E^{\textrm{QCF}}(t)\right] \\
        &  \propto \nonumber \frac{1}{\sqrt{2}}\left[\alpha^{\textrm{LO}} \e^{-\irm(\omega_{0}\,t+\Phi)} + \left(\frac{1 + g}{\sqrt{2\,g}}\,\alpha_{\Omega} \,\cos (\omega_{0}\,t  + \Omega\,t)\right.\right.\\
        &\p\quad \left.\left.-\frac{1 - g}{\sqrt{2\,g}}\,\alpha_\Omega\,\cos (\omega_{0}\,t  - \Omega\,t  - 2\,\phi)\right)\right]+ \textrm{c.c.}
\end{align}
\begin{align}
   E^{\textrm{HD2}} &= \nonumber \frac{1}{\sqrt{2}}\left[E^{\textrm{LO}} - E^{\textrm{QCF}}(t)\right] \\
        &  \propto \nonumber \frac{1}{\sqrt{2}}\left[\alpha^{\textrm{LO}} \e^{-\irm(\omega_{0}\,t+\Phi)} - \left(\frac{1 + g}{\sqrt{2\,g}}\,\alpha_\Omega \,\cos (\omega_{0}\,t  + \Omega\,t)\right.\right.\\
        &\p\quad \left.\left.-\frac{1 - g}{\sqrt{2\,g}}\,\alpha_\Omega\,\cos (\omega_{0}\,t  - \Omega\,t  - 2\,\phi)\right)\right]+
        \textrm{c.c.}\,.
\end{align}
\begin{figure}[t]
\centerline{\includegraphics[width=6.0cm,keepaspectratio]{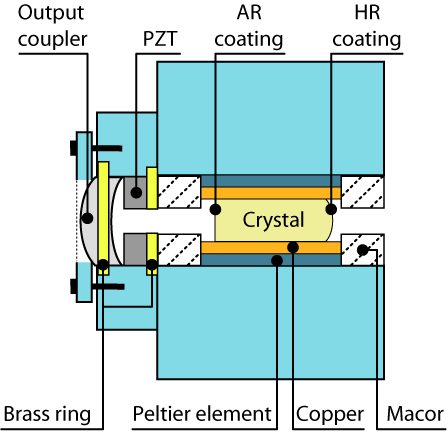}} 
\vspace{0mm} \caption{\coon. Cut through the squeezed-light
source. The hemilithic cavity is formed by the highly
reflection-coated crystal back surface and an outcoupling mirror.
The inside of the oven consists of two Peltier elements which are
used to actively stabilize the temperature of the crystal. The
thermal contact to the crystal is done via copper plates. To
thermally shield the crystal from outside, Macor blocks are used.
These have two small drillings on the optical axis for the laser
beam. The temperature sensor is embedded into the copper plates. A
stack made out of a brass ring, a PZT, the outcoupling mirror, and
a Viton ring is clamped together with an aluminum cap. This stack
is bolted onto the oven to have a rigid cavity.}
  \label{oven1}
\end{figure}
The difference current $I_{\textrm{diff}}$ of the induced
photocurrents $I^\textrm{HD1,2}=| E^{\textrm{HD1,2}}|^2$ is then
given by
\begin{eqnarray}
  I_{\textrm{diff}}&\propto&\nonumber\frac{4\,{\sqrt{2}}\,\alpha^{\textrm{LO}} \,\alpha_\Omega \,\cos (\Phi  + \omega_{0}\,t )}{{\sqrt{g}}}\\
  &\nonumber\p&\times\Big[ \left( 1 + g \right) \,\cos\left( \omega_{0}\,t  + \Omega\,t  \right)\\
  &\p&\quad-\left( 1 - g \right) \,\cos (\omega_{0}\,t  - \Omega\,t -2\,\phi)\Big]\\
  &\varpropto&\nonumber\frac{2\,{\sqrt{2}}\,\alpha^{\textrm{LO}} \,\alpha_\Omega\,\left( -1 + g \right)\,
    }{{\sqrt{g}}}\\&\p&\quad\times\cos (\Omega\,t + 2\,\phi  + \Phi
    )\,,
\label{I_diff}
\end{eqnarray}
where Eq.~(\ref{I_diff}) is the difference of the photocurrents
after low-pass filtering.  The demodulation of $I_{\textrm{diff}}$
with $\Omega$ and again low-pass filtering results in the error
signal $S\mathrm{_{err}^{QCF-LO}}$ for the relative phase $\Phi$
between the second-harmonic pump and the local oscillator
\begin{equation}
  S\mathrm{_{err}^{QCF-LO}}\varpropto\frac{\sqrt{2}\,\alpha^{\textrm{LO}}\,\alpha_\Omega\,\left( -1 + g \right)}{{\sqrt{g}}}\sin (2\,\phi  + \Phi)\,.
\label{Err_LO}
\end{equation}
This error signal depends not only on the relative phase between
the second-harmonic pump and the local oscillator $\Phi$, but also
on the squeezing angle $\phi$. However, it becomes clear that the
combination of both error signals according to
Eqs.\,(\ref{Err_AOM}) and (\ref{Err_LO}) enables full coherent
control of the squeezed vacuum generated by an OPO with respect to
the local oscillator of a downstream experiment.

\section{Experimental Setup and Results}

Figure~\ref{experiment} shows the schematic of our experimental
setup that was used to demonstrate coherently controlled broadband
vacuum squeezing. Altogether two independent, but phase locked
laser sources (lasers 1 and 2) were utilized. Both were monolithic
nonplanar neodymium-doped yttrium aluminum garnet (Nd:YAG) ring
laser of 2 and 1.2\,W single-mode output powers at 1064\,nm,
respectively. Approximately 1.4\,W of laser 1 was used to pump a
second-harmonic-generation (SHG) cavity. The design of the SHG
cavity was the same as for our OPO cavity. The two differed only
in the reflectivities of the outcoupling mirrors.

\begin{figure}[t]
\centerline{\includegraphics[width=8cm]{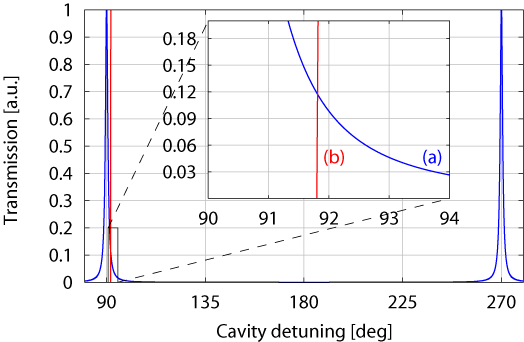}} 
\caption{\coon. {Theoretical OPO cavity transmission [curve (a)]
versus cavity detuning (peak normalized to unity). The cavity
linewidth is 29.8\,MHz. The 40\,MHz frequency offset from the
carrier frequency of the AOM --- producing the QCF --- is marked
with the plumb line [curve (b)].}}
  \label{AOM-lw}
\end{figure}

A detailed sketch of our SHG and OPO cavity design and mounting is
shown in Fig.~\ref{oven1}. We used a hemilithic layout which was
formed by an outcoupling mirror and a highly reflection-coated
crystal back surface. The oven was made of an aluminum surrounding
that houses the crystal and the Peltier elements and served as a
heat sink. The outcoupling mirror together with the piezoelectric
transducer (PZT) and a Viton ring formed a stack that was clamped
together with an aluminum cap. This stack was located in an
aluminum plate that was bolted onto the aluminum surrounding of
the oven. Macor blocks were used to thermally shield the crystal
from outside. The Macor blocks had two small holes on the optical
axis for the laser beam. A servo loop was used for active
temperature stabilization of the crystal. The required temperature
sensor was embedded into the copper plates and the servo feedback
was put on the peltier elements. The OPO as well as the SHG
crystal were made from 7\% doped MgO:LiNbO$_{3}$ and had the
dimensions 2.5$\times$5$\times$6.5\,mm$^3$. The curved back
surface of the crystals had a high-reflection coating
($R$=99.96\%) whereas the flat surface had an antireflection
coating ($R<$0.05\%) for both wavelengths. The cavities had a free
spectral range of approximately $4$\,GHz. The crystals were
mounted into the ovens in such a way that \emph{s}-polarized
fields could sense the nonlinearity

The SHG used an outcoupling mirror with power reflectivities of
$R_{\textrm{1064\,nm}}=92\%$ and $R_{\textrm{532\,nm}}<4\%$. The
cavity length was controlled using the Pound-Drever-Hall (PDH)
locking scheme with a phase modulation at a sideband frequency of
18\,MHz. The generated second-harmonic field had a power of up to
500\,mW.
\begin{figure}[t]
\centerline{\includegraphics[width=8.4cm]{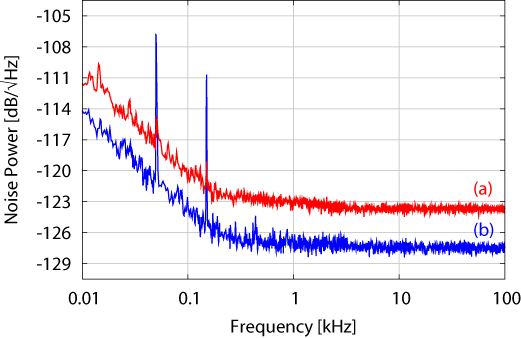}} 
\caption{\coon. Measured quantum noise spectra at sideband
frequencies $\Omega_s/2\pi$: (a) shot-noise and (b) squeezed noise
with $88$\,$\mu$W local oscillator power. All traces are pieced
together from five fast Fourier transform frequency windows:
10--50\,Hz, 50--200\,Hz, 200--800\,Hz, 800--3.2\,kHz, and
3.2--10\,kHz. Each point is the averaged rms value of 100, 100,
400, 400, and 800 measurements in the respective ranges. The RBWs
of the five windows were 250\,mHz, 1\,Hz, 2\,Hz, 4\,Hz, and
16\,Hz, respectively.}
  \label{ShotcombiSQZ}
\end{figure}

The OPO cavity utilized an outcoupling mirror with power
reflectivities of $R_{\textrm{1064\,nm}}=95.6\%$ and
$R_{\textrm{532\,nm}}=20\%$. This resulted in a linewidth of
28.9\,MHz at 1064nm (see Fig.~\ref{AOM-lw}). For the OPO or OPA
two different control loops for stabilizing the cavity length were
set up. The first cavity length control loop was used during
alignment of our experiment. It utilized a resonant
\emph{s}-polarized seed beam that carried phase modulation
sidebands at a frequency of 12\,MHz for a PDH locking scheme.  The
error signal could be generated using either the sum of the
homodyne detectors or an additional detector placed in
transmission of the OPA (see Fig.~\ref{experiment}). The latter
detector was also used to determine the frequency offset between
the \emph{s}- and \emph{p}-polarized laser beams inside the OPA or
OPO. The error signal was fed back to the PZT-mounted output
coupler.
The second cavity length control loop was used for the generation
of squeezed vacuum states at low frequencies, since the first
control loop introduced too much noise at low frequencies (see
Sec.~\ref{sec2}). This control was realized with a
\emph{p}-polarized field generated by the second monolithic
nonplanar Nd:YAG ring laser (laser 2). Due to the birefringence of
the MgO:LiNbO$_{3}$ crystal the $\textrm{TEM}_{00}\!$'s of the OPO
cavity for \emph{s}- and \emph{p}-polarization are not degenerate.
To ensure that  both polarizations resonate simultaneously in the
cavity we shifted the frequency of the \emph{p}-polarized field.
We determined the frequency shift to be about 1.4\,GHz. The
frequency offset was controlled via a phase-locking loop (PLL)
that could be operated from nearly DC up to 2\,GHz with a
bandwidth of several kilohertz. The error point of the PLL was fed
back to the PZT of the second laser. Phase modulation sidebands at
a frequency of 153.8\,MHz were imprinted onto the
\emph{p}-polarized field, which was then injected through the back
surface of the OPO crystal. The transmitted part was spatially
separated from the \emph{s}-polarized squeezed vacuum with a
polarizing beam splitter (PBS) and detected by the photodetector.
A PDH locking technique was used to generate an error signal which
was fed back to the PZT of the OPO cavity.

\begin{figure}[t]
\centerline{\includegraphics[width=8.4cm]{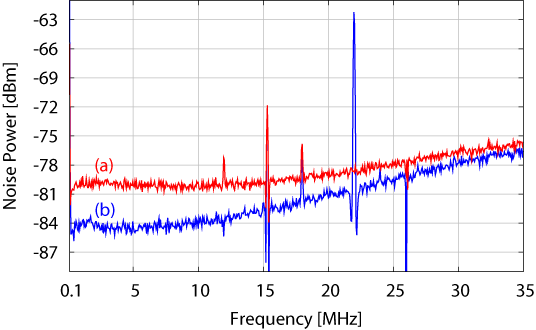}} 
\caption{\coon. Measured quantum noise spectra: (a) shot noise and
(b) squeezed noise with 8.9\,mW local oscillator power. The spikes
at 15.5, 22, and 26\,MHz are results from the dark noise
correction of modulation peaks due to other electronic fields.}
\label{SQZMHZ}
\end{figure}

Following our proposal described above we utilized a second
coherent but frequency-shifted control field for the phase control
of the squeezed vacuum field, detuned by $\Omega$ with respect to
the main carrier frequency ($\omega_{0}$, laser 1) by an AOM [see
Eq.~(\ref{quadratureamplitudes}) \& \ding{192} in
Figs.~\ref{experiment} and \ref{sidebands}]. The frequency of the
AOM was $\Omega/2\pi=40$\,MHz. This frequency-shifted
\emph{s}-polarized infrared QCF (440\,$\mu$W) was also injected
into the OPO cavity through the crystal's back surface. It
therefore had to be spatially overlapped with the
\emph{p}-polarized locking beam using a 50/50 beam splitter. To
eliminate technical noise below 1\,kHz in the homodyne spectra,
the zero order of the AOM had to be blocked carefully. If only
small fractions of this non-frequency-shifted field leaked into
the cavity, the squeezing spectrum was spoiled by the large
technical noise in the low-frequency regime.

Figure~\ref{AOM-lw} shows that only 11.5\% of the QCF was coupled
into the cavity. This 11.5\% interacted with the pump field inside
the cavity, its quadratures were parametrically amplified and
deamplified, and an additional sideband at $-40$\,MHz [see
Eq.~(\ref{E(t)})] was generated. This outgoing QCF from the OPO
then consisted of two sidebands, each separated by 40\,MHz from
the carrier frequency $\omega$ (see \ding{193} in
Figs.~\ref{experiment} and \ref{sidebands}). The error signal
could be obtained by detecting the outgoing QCF from the OPO and
demodulating the photocurrent at 80\,MHz as illustrated in the
sideband scheme in Figs.~\ref{experiment} and \ref{sidebands}. By
feeding back the error signal to a PZT-mounted mirror in the path
of the second-harmonic pump field, stable control of $\phi$ was
realized [see Eq.~(\ref{Err_AOM})]. The error signal for
controlling the homodyne angle $\Phi$ [see Eq.~(\ref{Err_LO})] was
derived at the homodyne detector. The difference of the two
photodiode currents was demodulated with a frequency of 40\,MHz.
The output of this servo loop was fed back to a PZT-mounted mirror
in the local oscillator path.

\begin{figure}[t]
\centerline{\includegraphics[width=8.4cm]{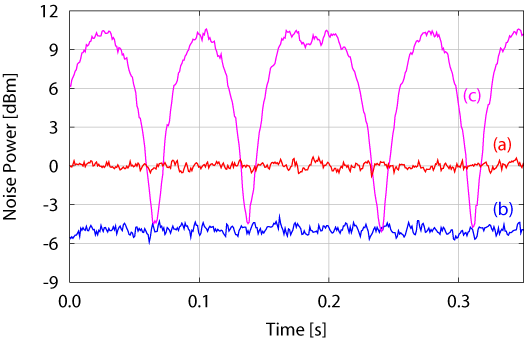}}  
\caption{\coon. Time series of shot noise [curve (a)], squeezed
noise with locked local oscillator phase [curve (b)] and squeezed
noise with scanned local oscillator phase [curve (c)] at
$\Omega_s/2\pi=5$\,MHz sideband frequency. A nonclassical noise
suppression of about $5.0 \pm 0.1$\,dB is demonstrated. The dark
noise has been subtracted.} \label{MACSQZ}
\end{figure}

The homodyne detector was built from a
\emph{p}-polarization-optimized 50/50 beam splitter and two
electronically and optically matched photodetectors based on
Epitaxx ETX500 photodiodes. The angular orientation of the
photodiodes was optimized to achieve the maximum power to optimize
the detection efficiency. We used two different pairs of matched
photodetectors: one was optimized for the low-frequency regime
whereas the other pair was optimized for high bandwidth. In all
measured spectra shown here the electronic noise of the detection
system was subtracted from the measured data.

The low-frequency-optimized homodyne photodiode pair permits the
measurement of the low-frequency spectrum of the OPO (see
Fig.~\ref{ShotcombiSQZ}) in the detection bandwidth of GW
interferometers and above up to 100\,kHz. For these low-frequency
measurements we used a nominal local oscillator power of
88\,$\mu$W. The resulting shot-noise limit of the homodyne
detection system is shown as curve (a) in Fig.~\ref{ShotcombiSQZ},
whereas the squeezed quantum noise is shown in curve (b). During
the measurement period of approximately 1.5 \,h the complete
experiment including the OPO with all its related control loops,
was controlled stably in all degrees of freedom. For the
measurements in Fig.~\ref{ShotcombiSQZ} a parametric gain of 10
was used, which was obtained using 60\,mW of the second-harmonic
pump field. The propagation losses of the squeezed vacuum were
dominated by the Faraday rotator passthrough efficiency of only
95\%. The mode matching efficiency  between the local oscillator
and the squeezed field was measured to be 94.3\%. Altogether this
allowed us to measure 4\,dB squeezing over the complete detection
band of ground-based GW interferometers.

\begin{figure}[]
\centerline{\includegraphics[width=8.0cm]{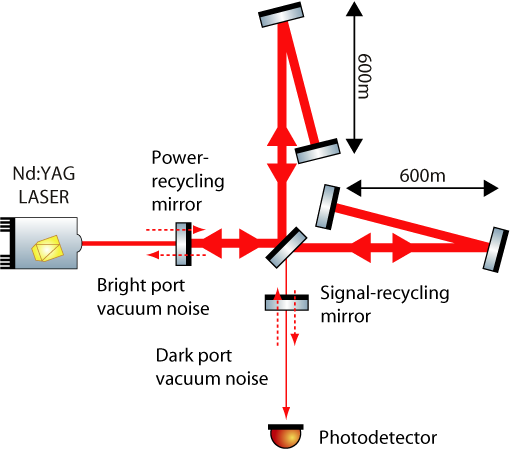}}  
\vspace{0mm} \caption{\coon. Simplified schematic of the
gravitational wave detector GEO\,600. The dashed arrows represent
the relevant vacuum fields entering and leaving the interferometer
if optical loss inside the interferometer can be neglected.
Generally, the squeezed vacuum states need to be injected into the
interferometer in such a way that they constructively interfere in
the dark signal output port for all relevant frequencies.}
  \label{GEO-sqz-input}
\end{figure}

The OPO squeezing spectrum from 100\,kHz to 35\,MHz is shown in
Fig.~\ref{SQZMHZ}. Curve (a) shows the shot-noise limit of the
homodyne detector, while the squeezing spectrum is shown in curve
(b). Both traces take the dark noise into account. These
measurements were done with the high-bandwidth-optimized homodyne
detector using a local oscillator power of 8.9\,mW. The resulting
shot noise showed a nonwhite behavior above 10\,MHz. This stemmed
from small deviations in the transfer functions of the two
homodyne photodetectors used. One can see that up to 10\,MHz we
observed at least 4\,dB of squeezing, peaking around 5 MHz with up
to 4.95\,dB of squeezing (see Fig.~\ref{MACSQZ}). At higher
frequencies the squeezing degraded to approximately 1\,dB due to
the linewidth of the OPO cavity. Lowering the finesse of the OPO
would open up the high-frequency regime for a better squeezing
performance, but higher pump powers would be needed to produce the
same amount of squeezing.

The two individual measurements show that we have produced 4\,dB
of squeezing over more than six decades from 10\,Hz up to 10\,MHz.

\section{Application to Gravitational Wave Detectors}

In \cite{VCHFDS06} a broadband squeezed vacuum field was applied
to a simple Michelson interferometer. A nonclassical
signal-to-shot-noise improvement was observed using balanced
homodyne detection. However, real gravitational wave Michelson
interferometers are much more complex. Here we discuss two aspects
that are important when coherently controlled broadband squeezed
vacuum fields are applied to signal-recycled gravitational wave
detectors with heterodyne readout.

In an application to a gravitational wave detector, the quadrature
control field, here shifted in frequency by $\Omega/2\pi=40$\,MHz,
will enter the interferometer from the dark port together with the
squeezed field \cite{Cav81,SHSD04}. The best choice for the QCF
frequency is such that it is offresonant with respect to the
signal-recycling cavity (SRC). In this case the QCF is basically
reflected from the SRC, which minimizes possible disturbances to
other interferometer control loops. A generally rather important
interferometer control field stabilizes the Michelson
interferometer on a defined differential arm length to provide the
desired dark signal port condition. In GEO\,600 this control field
operates at a sideband frequency of 14.9\,MHz and is
photoelectrically detected in the dark port. Obviously, the QCF
frequency $\Omega$ should also provide a sufficiently large offset
from that frequency of 14.9\,MHz. Our choice of
$\Omega/2\pi=40$\,MHz is therefore rather practical in the case of
GEO\,600.

Another aspect is the compatibility of the squeezed vacuum field
demonstrated here and the heterodyne detection scheme that is
currently used in all gravitational wave detectors, namely
GEO\,600~\cite{geo04,geo06}, LIGO~\cite{LIGO,geo04},
TAMA\,300~\cite{TAMA}, and VIRGO~\cite{VIRGO04}. Gea-Banacloche
and Leuchs \cite{GLe87} have shown that an interferometer with
heterodyne readout requires squeezing in the band of expected
gravitational wave signals ($\Omega_s$) and also squeezing around
twice the heterodyne frequency ($2\omega_m\!\pm\Omega_s$). This is
because the noise at $\omega_m\!\pm\Omega_s$ contains vacuum noise
contributions from $2\omega_m\!\pm\Omega_s$. It was later shown
that these results are still valid for detuned signal-recycled
interferometers \cite{CDRGBM98}.

The realization that broadband squeezing up to
$2\omega_m\!\pm\Omega_s$ is needed to gain the full sensitivity
enhancement from squeezing leads to the question as to where one
should inject the squeezed vacuum field.
Figure~\ref{GEO-sqz-input} shows a schematic of GEO\,600. In
normal operation the interferometer behaves like an almost perfect
mirror for the carrier light entering from the bright port and for
vacuum fluctuations entering from the dark port. Due to a small,
but macroscopic, difference of the two arm lengths, which is
needed for the Schnupp modulation \cite{Schnupp88}, the
reflectivity of this mirror is frequency dependent. Thus, the
amplitude reflectivity $r(\omega)$ changes at different sideband
frequencies, relative to the carrier frequency. The decision as to
which interferometer port the squeezed states will be injected
into, is therefore dependent on the sideband frequency.

We can distinguish between three different cases. First, the
interferometer has high reflectivities in the two interesting
frequency regions: $r(\Omega_s)\thickapprox
r(2\omega_m\!\pm\Omega_s)\thickapprox 1$. If so, we can inject the
squeezed states at all frequencies from the dark port. These will
then be perfectly reflected and we obtain the full improvement
from the squeezing.

The second case is that we have a high reflectivity at the signal
frequency $\Omega_s$ but high transmission at frequency
$2\omega_m\!\pm\Omega_s$, $r(\Omega_s)\thickapprox 1$; $
r(2\omega_m\!\pm\Omega_s)\thickapprox 0$. If we still injected all
the squeezed states from the dark port, we would lose the
squeezing at $2\omega_m\!\pm\Omega_s$, which results in a less
sensitive interferometer. To solve this problem one might split
the squeezed-light field in frequency space e.g. use a filter
cavity. We then obtain a field that carries the low-frequency
squeezing around $\Omega_s$ and a second field that is squeezed
around frequencies $2\omega_m\!\pm\Omega_s$. The low-frequency
squeezed field is injected through the dark port, whereas the
second one has to be injected through the bright port together
with the carrier field. This will then give the optimal
performance increase one expects from using squeezed light.
Instead of using a filter cavity, one might employ two independent
sources of squeezed states with optimized nonclassical noise
suppression in the audio and rf bands, respectively.
The third case is that the interferometer reflectivity is still
high at the signal frequencies, $r(\Omega_s)\thickapprox 1$, but
has an intermediate reflectivity, $0<r(2\omega_m\!\pm\Omega_s)<1$,
near twice the modulation frequency. In this case the power
fraction of $r(2\omega_m\!\pm\Omega_s)^2$ has to be tapped off the
high-frequency part of the squeezed field. This fraction has to be
injected into the interferometer's bright port, whereas the
remaining fraction is injected into the dark port, together with
100\% of the low-frequency part of the squeezed field. In this way
the squeezed field at high frequencies senses a Mach-Zehnder-type
configuration and constructively interferes in the
interferometer's (dark) signal port. Again the broadband squeezed
field is optimally employed for a nonclassical sensitivity
improvement of a gravitational wave detector with heterodyne
readout. We note that in all these three cases one has to use
frequency-dependent squeezed fields to compensate the phase shifts
from the reflection at a detuned cavity
\cite{HCCFVDS03,CVHFLDS05}.

GEO\,600 currently uses a heterodyne frequency of 14.9\,MHz. The
reflectivity of the interferometer at this frequency is
approximately 96\% in power. Consider now a broadband vacuum
squeezed field of 6\,dB nonclassical noise suppression from 10\,Hz
up to 30\,MHz injected into GEO\,600's dark port. Neglecting
optical losses inside the interferometer, the squeezed states at
gravitational wave signal frequencies are perfectly reflected.
Those at twice the heterodyne frequency sense 4\% loss, and their
nonclassical noise suppression degrades from 6 to about 5.5\,dB,
which is still a useful value. Hence the injection of the complete
broadband vacuum squeezed field into the dark port seems to be a
reasonable approach in the case of GEO\,600.

\section{Conclusion}

We have reported on a control scheme for phase locking of squeezed
vacuum fields, generated by optical parametric oscillation, to a
local oscillator of a downstream experiment or of a homodyne
detector. Our scheme utilized two frequency-shifted control fields
that allowed us to control the length of the OPO cavity as well as
the angle of the squeezed-field quadrature. This control scheme
allowed stable generation and observation of broadband squeezed
fields covering more than six decades from 10\,Hz to about
35\,MHz. We discussed the application of our control scheme and
the broadband squeezed field generated for GEO\,600, as an example
for a large-scale gravitational wave detector. We found that such
a squeezed field injected into the signal dark port
can improve GEO\,600's sensitivity beyond its shot-noise limit, even if the current heterodyne readout is used.\\

\section{Acknowledgements}
We thank Alexander Franzen, Boris Hage, and Jan Harms for fruitful
discussions.\\

This work has been supported by the Deut\-sche
For\-schungs\-ge\-mein\-schaft and is part of the
Son\-der\-for\-schungs\-be\-reich 407.

\appendix


\begin{thebibliography}{12}

\bibitem{Wal83} D.~F.~Walls, Nature {\bf 306}, 141 (1983). 

\bibitem{Cav81} C.~M.~Caves, Phys. Rev. D {\bf 23}, 1693 (1981).

\bibitem{Thorne87} K.S.~Thorne, in {\it 300 Years of Gravitation}, edited by S.W.~Hawking and W.~Isreal (Cambridge University Press, Cambridge, England, 1987), pp.~330--458.

\bibitem{Unruh82} W.~G.~Unruh, in  {\it Quantum Optics, Experimental Gravitation, and Measurement Theory}, edited by P.~Meystre and M.~O.~Scully (Plenum, New York, 1983), pp 647--660.

\bibitem{KLMTV01} H.~J.~Kimble, Y.~Levin, A.~B.~Matsko, K.~S.~Thorne and S.~P.~Vyatchanin, Phys. Rev. D {\bf 65}, 022002 (2001).

\bibitem{HCCFVDS03} J.~Harms, Y.~Chen, S.~Chelkowski, A.~Franzen, H.~Vahl\-bruch, K.~Danzmann, and R.~Schnabel, Phys. Rev. D {\bf 68}, 042001 (2003).

\bibitem{Mee88} B.~J.~Meers, Phys. Rev. D {\bf 38}, 2317 (1988).

\bibitem{SHYMV85} R.~E.~Slusher, L. W. Hollberg, B.~Yurke, J. C. Mertz, and J. F. Valley, Phys. Rev. Lett. \textbf{55}, 2409 (1985).

\bibitem{Shoemaeker03} D.~Shoemaker, Classical Quantum Gravity {\bf 20}, S11 (2003).

\bibitem{BGV99} V.~B.~Braginski, M.~L.~Gorodetsky, and S.~P.~Vyatchanin,
Phys. Lett. A {\bf 264}, 1 (1999).

\bibitem{GLe87} J.~Gea-Banacloche and G.~Leuchs, J.~Mod.~Opt. {\bf 34}, 793 (1987).

\bibitem{CVHFLDS05} S.~Chelkowski, H.~Vahlbruch, B.~Hage, A.~Franzen, N.~Lastzka, K. Danzmann, and R. Schnabel, Phys. Rev. A {\bf 71}, 013806 (2005).

\bibitem{VCHFDS05} H.~Vahlbruch, S.~Chelkowski, B.~Hage, A.~Franzen, K. Danzmann, and R. Schnabel, Phys. Rev. Lett. {\bf 95}, 211102 (2005).

\bibitem{VCHFDS06CQG} H.~Vahlbruch, S.~Chelkowski, B.~Hage, A.~Franzen, K.~Danzmann, R.~Schnabel, Class. Quantum Grav. {\bf 23} S251 (2006).

\bibitem{MGBWGML04} K.~McKenzie, N.~Grosse, W.~P.~Bowen, S.~E.~Whitcomb, M.~B.~Gray, D.~E.~McClelland, and P.~K.~Lam, Phys. Rev. Lett. {\bf 93}, 161105 (2004).

\bibitem{MMGLGGMM05} K. McKenzie, E. E. Mikhailov, K. Goda, P. K. Lam, N. Grosse, M. B. Gray, N. Mavalvala, and D. E. McClelland, J. Opt. B {\bf 7}, S421 (2005).

\bibitem{VCHFDS06} H.~Vahlbruch, S.~Chelkowski, B.~Hage, A.~Franzen, K. Danzmann, and R. Schnabel, Phys. Rev. Lett. {\bf 97}, 011101 (2006).

\bibitem{CSc85} C. M. Caves and B. L.Schumaker, Phys. Rev. A {\bf 31}, 3068 (1985)


\bibitem{SHSD04} R. Schnabel, J. Harms, K. A. Strain, and K. Danzmann, Class. Quantum Grav. \textbf{21}, S1045 (2004).

\bibitem{geo06} S.~Hild (for the LIGO Scientific Collaboration), Class. Quantum Grav. {\bf 23}, S643 (2006).

\bibitem{geo04} B.~Abbott {\it et al.}, Nuclear Instruments and Methods in Physics
Research Section A \textbf{517}, 154 (2004).

\bibitem{LIGO} A.~Abramovici {\it et al.}, Science {\bf 256}, 325 (1992).

\bibitem{TAMA} M.~Ando {\it et al.}, Phys.~Rev.~Lett. {\bf 86}, 3950 (2001).

\bibitem{VIRGO04}F.~Acernese {\it et al.}, Class. Quantum Grav. {\bf 21}, S709 (2004).

\bibitem{CDRGBM98} V.~Chickarmane, S.~V.~Dhurandhar, T.~C.~Ralph, M. Gray, H.-A.~Bachor, and D.~E.~McClelland, Phys. Rev. A {\bf 57}, 3898 (1998).

\bibitem{Schnupp88} L. Schnupp: talk at the European Collaboration Meeting on Interferometric
Detection of Gravitational Waves (Sorrento, 1988).

\end{thebibliography}
\end{document}